\documentclass[sigconf, authorversion]{acmart}
\usepackage{booktabs}
\usepackage{array}
\usepackage{xcolor}
\usepackage{stfloats}

\newcolumntype{Y}{>{\raggedright\arraybackslash}X} 

\AtBeginDocument{%
  }

\copyrightyear{2026}
\acmYear{2026}
\setcopyright{cc}
\setcctype{by}
\acmConference[DIS '26]{Designing Interactive Systems Conference}{June 13--17, 2026}{Singapore, Singapore}
\acmBooktitle{Designing Interactive Systems Conference (DIS '26), June 13--17, 2026, Singapore, Singapore}
\acmDOI{10.1145/3800645.3812947}
\acmISBN{979-8-4007-2563-0/2026/06}

\begin{document}

\title{Speculating the Impacts of Mediated Social Touch Technology}


\author{Russian (Ruo-Xuan) Wu}
\orcid{0009-0003-5806-8382}
\affiliation{
  \institution{Design Lab, School of Architecture, Design and Planning\\ The University of Sydney}
  \city{Sydney}
  \state{NSW}
  \country{Australia}}
\email{russian.wu@sydney.edu.au}

\author{Tim Moesgen}
\orcid{0000-0003-2016-2261}
\affiliation{%
  \institution{Department of Information and Communications Engineering\\ Aalto University}
  \city{Espoo}
  \country{Finland}}
\email{tim.moesgen@aalto.fi}

\author{Myung Jin (MJ) Kim}
\orcid{0000-0001-9970-4056}
\affiliation{%
  \institution{Human Sensory Augmentation Research Section\\ Electronics and Telecommunications Research Institute}
  \city{Daejeon}
  \country{Republic of Korea}}
\email{mj@etri.re.kr}

\author{Xinyan Yu}
\orcid{0000-0001-8299-3381}
\affiliation{%
  \institution{Design Lab, School of Architecture, Design and Planning\\ The University of Sydney}
  \city{Sydney}
  \state{NSW}
  \country{Australia}}
\email{xinyan.yu@sydney.edu.au}

\author{Naoki Kameyama}
\orcid{0009-0009-6858-347X}
\affiliation{%
  \institution{Graduate School of Design\\ Kyushu University}
  \city{Fukuoka}
  \country{Japan}}
\email{kameyama3250@gmail.com}

\author{Anusha Withana}
\orcid{0000-0001-6587-1278}
\affiliation{%
  \institution{School of Computer Science\\ Sydney Nano Institute \\ The University of Sydney}
  \city{Sydney}
  \state{NSW}
  \country{Australia}}
\email{anusha.withana@sydney.edu.au}

\author{Marius Hoggenmueller}
\orcid{0000-0002-8893-5729}
\affiliation{%
  \institution{Design Lab, School of Architecture, Design and Planning\\ The University of Sydney}
  \city{Sydney}
  \state{NSW}
  \country{Australia}}
\email{marius.hoggenmueller@sydney.edu.au}

\author{Luke Hespanhol}
\orcid{0000-0003-0839-481X}
\affiliation{%
  \institution{Design Lab, School of Architecture, Design and Planning\\ The University of Sydney}
  \city{Sydney}
  \state{NSW}
  \country{Australia}}
\email{luke.hespanhol@sydney.edu.au}

\renewcommand{\shortauthors}{Wu et al.}

\begin{abstract}
  With growing research on haptic interfaces, Mediated Social Touch (MST) technologies offer the potential to record, synthesise, and reproduce (RSR) touch experiences across space and time, enabling, for instance, a hug from afar and from the past. Although much of the existing research highlights the direct benefits of these systems, such as reducing loneliness and providing emotional support, little attention has been paid to their \textcolor{black}{broader sociotechnical impacts}. To address this gap, we used the \textit{Future Ripples} method to speculate on possible effects of MST. We conducted three workshops with 24 participants, including potential users, domain experts, and haptics researchers. Throughout these sessions, participants collectively envisioned possible future scenarios, alongside opportunities and threats, and proposed actionable responses. Our qualitative analysis organised these insights into four themes and three distinctive challenges. \textcolor{black}{These findings offer haptics researchers intervention points across the RSR pipeline to inform MST design, alongside methodological insights from applying Future Ripples to MST technology.}
\end{abstract}

\begin{abstract}
  With growing research on haptic interfaces, Mediated Social Touch (MST) technologies offer the potential to record, synthesise, and reproduce (RSR) touch experiences across space and time, enabling, for instance, a hug from afar and from the past. Although much of the existing research highlights the direct benefits of these systems, such as reducing loneliness and providing emotional support, little attention has been paid to their broader sociotechnical impacts. To address this gap, we used the Future Ripples method to speculate on possible effects of MST. We conducted three workshops with 24 participants, including potential users, domain experts, and haptics researchers. Throughout these sessions, participants collectively envisioned possible future scenarios, alongside opportunities and threats, and proposed actionable responses. Our qualitative analysis organised these insights into four themes and three distinctive challenges. These findings offer haptics researchers intervention points across the RSR pipeline to inform MST design, alongside methodological insights from applying Future Ripples to MST technology.
\end{abstract}




\begin{teaserfigure}
  \includegraphics[width=\textwidth]{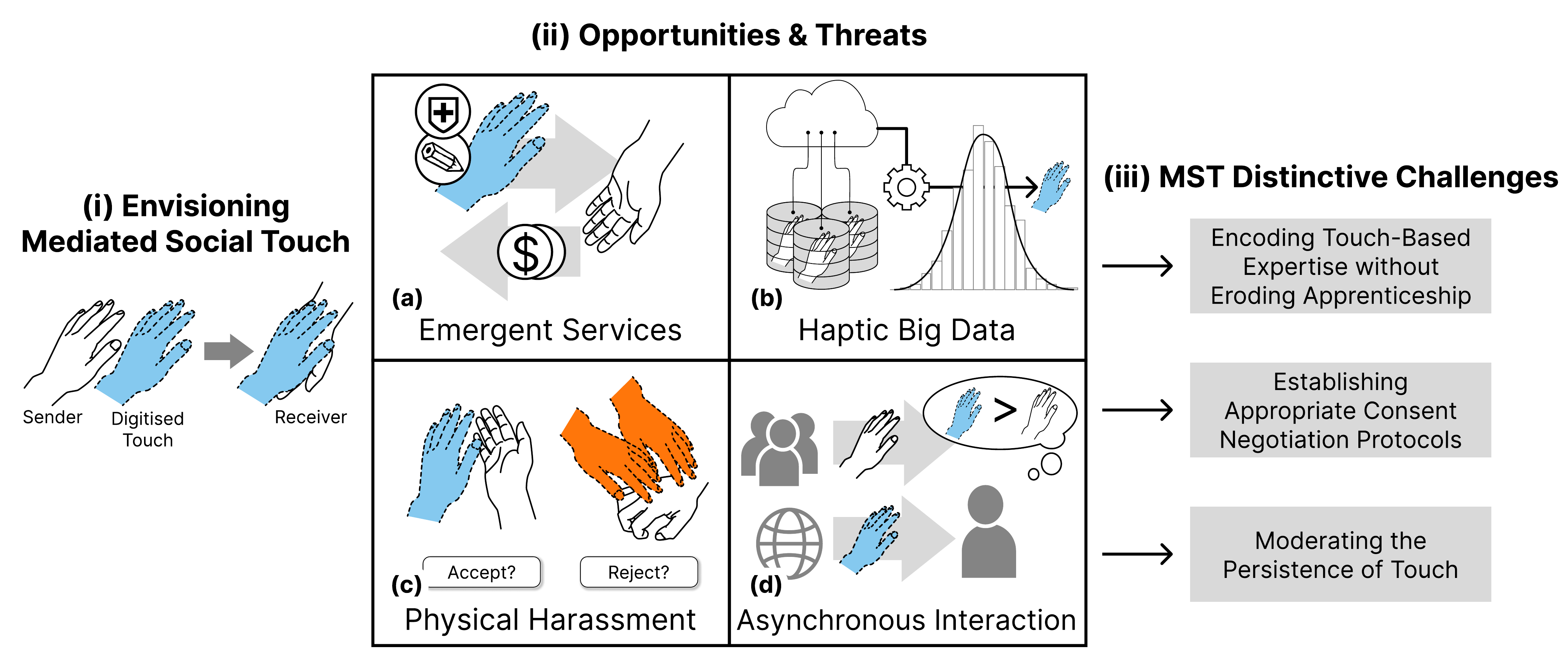}
  \caption{In our work we present: (i) Future Ripples workshops on Mediated Social Touch (MST), (ii) four identified themes of opportunities and threats (a-d), and (iii) three distinctive challenges introduced by MST mapped to an adopted Recording–Synthesising–Reproducing (RSR) pipeline serving as intervention points in the MST design process.}
  \Description{The figure provides an overview of the paper’s conceptual structure. It visually summarises Mediated Social Touch (MST) as the central concept and situates it alongside three major analytical components: potential opportunities and threats, distinctive challenges, and the Recording–Synthesising–Reproducing (RSR) pipeline. Icons and arrows illustrate how embodied touch becomes digitised, how this enables new services, and how technical, ethical, and social issues emerge across the pipeline. The figure functions as a visual roadmap of the paper. Note: Description generated with ChatGPT.}
  \label{fig:teaser}
\end{teaserfigure}


\maketitle

\section{Introduction}
Technologies have allowed people to communicate and connect emotionally in numerous ways, including via telephone, television, and online video calls. Notwithstanding these advancements, one essential modality of human connection remains largely unsupported: touch. This absence became especially apparent during the COVID-19 pandemic, when physical distancing measures and travel restrictions intensified the importance of touch in remote communication, particularly as video calling became the dominant mode of maintaining emotional connections~\cite{jewitt_2021_manifesto}. Researchers and designers have thus increasingly explored technologies that use haptic feedback to stimulate the sense of touch~\cite{haans_mediated_2006}. Mediated Social Touch (MST) interfaces, accordingly, convey interpersonal touch through haptic technologies, enabling emotional bonding, remote education, training simulations, and therapeutic interventions~\cite{van_2015_SocialTouchinHCI}.

With MST, touch sensations such as pressure profiles can be recorded, edited and replayed on demand, regardless of time and place~\cite{Taguchi_2023_Multichannel}. For example, one might send a hug to a partner in a long-distance relationship~\cite{wu_2025_HuggingSuit}, learn piano techniques alongside an expert through guided haptic feedback~\cite{luo_adaptive_2024}, or even receive rehabilitation guidance delivered through the instructive motion of a therapist~\cite{Vianello_Exoskeleton_2024}. Nevertheless, such visions often prioritise technical feasibility and immediate benefit while giving limited consideration to the broader socio-technical implications. In particular, their long-term societal impacts remain underexplored.

In this stance, HCI researchers are calling for more research on the implications of \textit{digital social touch} and are encouraging designs with a stronger social and sensory focus ~\cite{jewitt_2021_manifesto}. This call was reinforced more recently by \citet{van_erp_editorial_2023}, who echoed concerns that MST is in crisis and briefly noted challenges relating to agency, control, and consent, mapping out opportunities for haptic researchers and designers. 
While such calls have drawn attention to the socio-technical implications of MST, two gaps remain. First, existing discussions have primarily centred on the perspectives of haptic researchers and designers~\cite{jewitt_2021_manifesto, barbareschi_moving_2025}. Yet, as MST moves closer to everyday use, the voices of potential users and domain experts from fields such as ethics, psychology, and law that may be affected by the widespread adoption of MST become equally critical for surfacing a broader range of opportunities and threats.
Second, even if we identify a range of opportunities and threats, it remains unclear which of these are distinctive to MST rather than common to emerging technologies in general. Touch differs fundamentally from other sensory modalities: it is inherently embodied, intimate, and deeply embedded in social and cultural norms~\cite{jewitt_2019_DigitalTouchEthics, Cekaite_InteractionalApproachTouch_2020}. As \citet{jewitt_2021_manifesto} urged, there is a lack of social visioning that accounts for such unique properties of touch. Understanding what is distinctive about MST is essential for generating actionable insights helpful for researchers and designers, rather than merely applying general frameworks. To address these gaps, we pose two research questions:

\begin{itemize}
    \item \emph{RQ1 What opportunities and threats emerge when potential users, domain experts, and haptic researchers and designers speculate about MST?}
    \item \emph{RQ2 What distinctive challenges can arise from widespread adoption of MST, and how can researchers address these challenges?}
\end{itemize}

\noindent To this end, we used the Future Ripples method~\cite{epp_reinventing_2022}, which materialises possible futures as metaphorical ``ripples'' or waves to help participants visualise and explore interconnected cause–effect relationships as MST evolves over time. We conducted three workshop iterations with a total of 24 participants: the first with 6 potential users; the second with 6 domain experts in biomedicine, law, psychology, ethics, dance, and design; and the third with 12 hapticians. Each workshop consisted of three structured activities that guided participants to envision possible futures of MST, identify associated opportunities and threats, and propose responses to the identified challenges. 

Specifically, five groups of participants created ripple maps for five MST applications: medical care, social interaction, education and training, emotional support, and online collaboration. From the analysis of those maps, four themes emerged that capture both opportunities and threats of MST: (1) Digitised Touch Catalyses Emergent Services, (2) Haptic Big Data Enables Data-Driven MST, (3) Remote Touch Complicates Consent and Exposes Users to Physical Harassment, and (4) Asynchronous Touch Reshapes Social Touch Practices. Participants also proposed actionable responses to address some of these threats. Building on these themes and responses, we further extracted three distinctive MST challenges: (1) Encoding Touch-Based Expertise without Eroding Apprenticeship, (2) Establishing Appropriate Consent Negotiation Protocols, and (3) Moderating the Persistence of Touch. We then adopted the recording–synthesising–reproducing (RSR) pipeline as an analytical lens~\cite{erk_2015_effects,minamizawa_techtile_2012,Taguchi_2023_Multichannel} \textcolor{black}{to examine these challenges across different stages of the MST design process.}


\textcolor{black}{
Motivated by the calls for more speculation about MST, we contribute to the design of haptic technologies in two ways:
\begin{enumerate}
    \item \textbf{Empirically,} we concretise potential futures of MST technology through five "what-ifs" and their consequences speculated by users and experts. From the resulting discussions, we identify four themes of opportunities and threats, and further extract three distinctive challenges posed by MST. We map these challenges onto the RSR pipeline to identify where they emerge and where mitigation strategies can be implemented within the MST design process.
    \item \textbf{Reflectively,} we share methodological insights from applying the Future Ripples method to speculate on MST as an emerging technology domain, discussing how facilitation choices such as prompt framing, group composition, and material setup shape the scope and direction of speculative outcomes.
\end{enumerate}
}

\section{Related Work}
\subsection{Defining Mediated Social Touch}

Social touch refers to any physical interaction involving direct and real-time contact between two or more individuals who are physically co-present in the same space~\cite{huisman_2017_SocialTouchTechnology, haans_mediated_2006, gallace_2010_InterpersonalTouch}. Following this definition, social touch encompasses not only intentional physical contact for social purposes but also incidental or functional touch, such as a punch in a sports match, a palpation in a medical examination, or an accidental brush of hands when receiving a coffee from a barista~\cite{van_2015_SocialTouchinHCI}. It plays a central role in human bonding, emotional regulation~\cite{bonanni_2006_TapTap, Jensen_SuspenderMender_2024}, and the development of trust and intimacy~\cite{suvilehto_2015_topography}.

MST builds upon the use of haptic technologies to transmit social touch between individuals who are not physically co-present. In this study, we adopt a broader interpretation of MST as ``the ability of one actor to touch another over both distance and time through haptic technology''~\cite{huisman_2017_SocialTouchTechnology, haans_mediated_2006, van_2015_SocialTouchinHCI}. For the past two decades, MST technologies have been explored in various forms, from wearable devices like YourGloves and TapTap~\cite{bonanni_taptap_2006, gooch_yourgloves_2012}, to more immersive garments such as the Huggy Pajama and the pneumatically actuated Hugging Suit~\cite{wu_2025_HuggingSuit,teh_huggy_2008}. These systems aim to support affective communication and bridge physical separation by delivering haptic sensations in diverse approaches and techniques. MST technologies are often implemented through three phases~\cite{Taguchi_2023_Multichannel, minamizawa_techtile_2012, erk_2015_effects}: (1) recording, capturing natural haptic sensations as they occur in the real world; (2) editing, adjusting properties of the recorded haptic data (e.g., timing, intensity, etc.); and (3) replaying, delivering the recorded sensations through haptic devices. 

However, the majority of the current literature remains focused on technology and increasing the fidelity of haptic simulation \textcolor{black}{often characterised as the "replica red herring" \cite{fairhurst_functional_2023, jewitt_2021_manifesto}. While a growing body of work has begun to raise ethical and social concerns \cite{jewitt_2019_DigitalTouchEthics, van_erp_editorial_2023, barbareschi_moving_2025}, there is still limited discussion on what happens when intimate, co-located touch is transformed into a technology-mediated experience that are detached from their original spatial, temporal, and contextual grounding. In the following section, we review methods for anticipating such consequences and the emerging speculative work on MST.}

\begin{figure*}[h]
    \centering
    \includegraphics[width=1\linewidth]
    {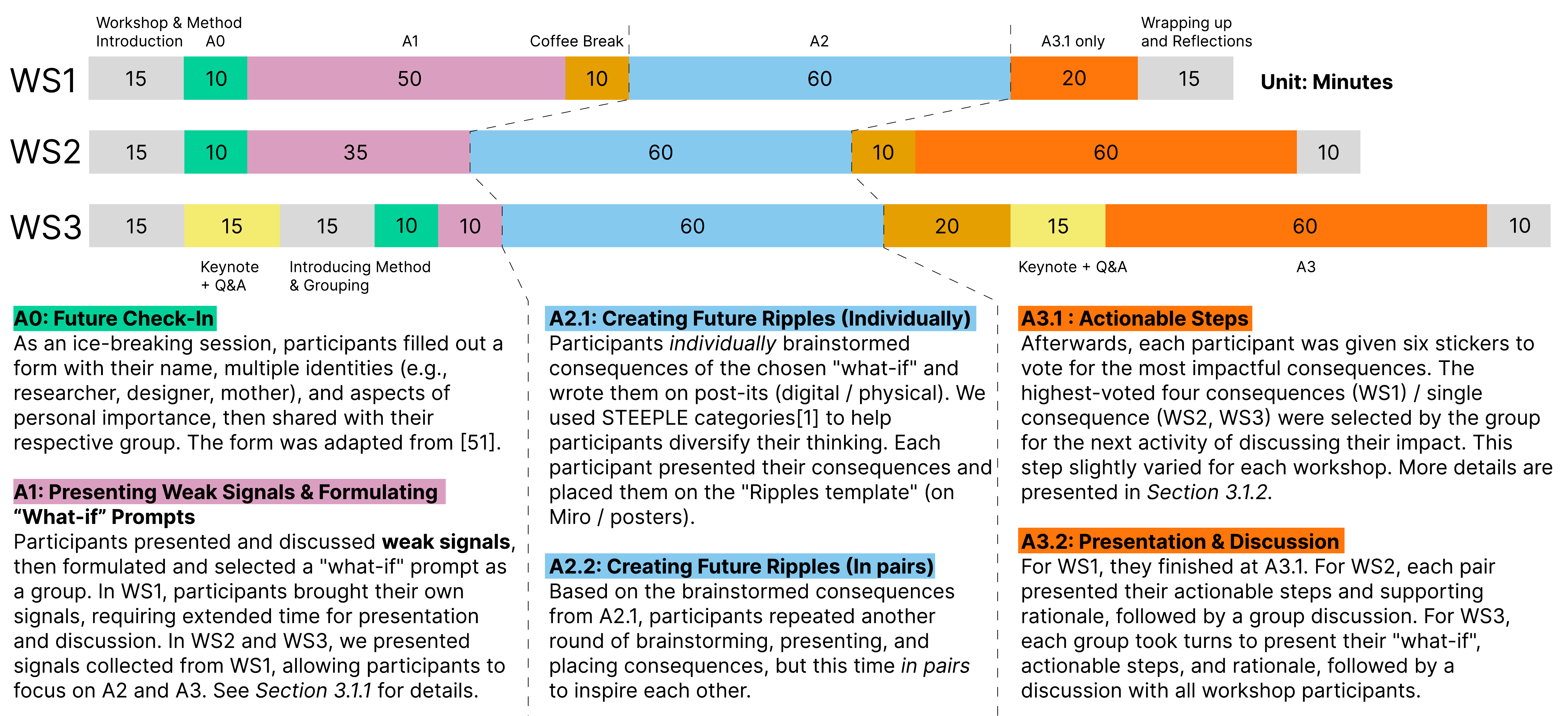}
    \caption{Workshop Activities Details. In this diagram, we visualised the different focuses of each workshop by presenting the time allocations and weightings for each activity. Workshop 1 (WS1) allocated more time to Activity 1 (A1) to broadly collect weak signals from their everyday lives. Workshop 2 and 3 (WS2 and WS3) allocated more time for Activity 3 (A3) to allow domain and haptic experts to propose potential solutions to the speculative consequences.}
    \label{tab:WorkshopActivities}
    \Description{Figure 2 outlines the sequence of workshop activities. The process began with A0 Future Check-in, an ice-breaker where participants completed a form about their identities and priorities and shared within their groups. In A1 Choosing a Pebble, participants presented weak signals and selected a “what-if” scenario as the basis for later exercises. Activity A2.1 involved participants individually brainstorming consequences of the chosen scenario and writing them on post-its, while A2.2 repeated the process in pairs to inspire further ideas. In A3.1 Actionable Steps, participants voted on the most impactful consequences, prioritised them, and analysed them using mapping tools. Finally, A3.2 Presentation had participants or groups share their selected “what-if,” proposed steps, and rationale, with small variations between workshops. Together, the table captures how the workshops moved from individual reflection to group discussion, prioritisation, and presentation.}
\end{figure*}

\subsection{Anticipating the Impacts of Disruptive Technologies}

In HCI, there have been amplified calls for speculating or ``anticipating'' unintended consequences of disruptive technologies \cite{jewitt_2019_DigitalTouchEthics, sturdee_consequences_2021, hecht_its_2021}. \textcolor{black}{A recent review of HCI's future orientation found that futuring in the field remains largely fleeting and techno-centric: most publications envision futures only as brief extrapolations of current technology, with limited attention to long-term, cascading consequences across social, ethical, and political dimensions~\cite{sanchez_lets_2025}.} Further, empirical studies show that even expert researchers struggle to foresee broader social impacts, motivating the development of new methods for anticipating side effects \cite{do_thats_2023}. Unintended consequences of technologies are future outcomes that are unforeseen and undesired, emerging when innovations interact with entangled social, economic, and environmental systems in unexpected ways~\cite{parvin_unintended_2020}. This can manifest as harmful ripple effects, inequities, or misuse, stemming from design blind spots, limited stakeholder inclusion, or short-term thinking. Anticipating such consequences is therefore critical in mitigating risks before they become embedded in everyday life. Next, we present approaches, methods and frameworks increasingly used by technology researchers to explore the impacts of disruptive technologies through speculation. 

Future-oriented approaches and methods such as speculative and critical design \cite{coombs_speculative_2018}, design fiction \cite{carta_design_2022}, and experiential futures \cite{candy_design_2019} have received growing interest from researchers and designers. Such methods advocate for critically reflecting on the impact of emerging technologies on our daily lives and discussing alternative futures that withstand the status-quo, using narratives, props, and situated experiences.

\textcolor{black}{Alongside these approaches, Value-Sensitive Design (VSD) offers a theoretically grounded framework for integrating human values throughout the technology design process~\cite{friedman_value_2008}. VSD has been applied across domains, from privacy-aware systems \cite{xu_value_2012} to equitable artificial intelligence (AI) \cite{liao_enabling_2019}. Complementing VSD at a broader level, the Responsible Innovation (RI) framework foregrounds four commitments, namely anticipation, reflexivity, inclusivity, and responsiveness, to guide the ethical governance of emerging technologies \cite{stilgoe_developing_2017}. In the haptics domain, RI has been applied to touchless mid-air haptics through expert workshops that generated social interaction applications through speculative exploration, with the resulting ideas analysed through RI's four dimensions \cite{cornelio_responsible_2023}. This analysis revealed that experts tend to assume benefits without examining underlying social values, imagine narrow stakeholder groups, and frame risks primarily as technical challenges. The study called for haptics researchers to embed more anticipatory thinking, for example, surfaced through "what-if" questioning, into their research practices.}

\textcolor{black}{While VSD and RI provide valuable commitments and principles, Future Ripples \cite{epp_reinventing_2022} -- the method employed in our paper -- offers a structured workshop technique that enables participants to collectively trace cascading consequences of a "what-if" scenario. It does so by scanning the present for weak signals and trends, and then iteratively mapping first-, second-, and third-order ripples across socio-technical layers. Where RI has primarily been used as an analytical framework to evaluate expert-generated ideas post-workshop \cite{cornelio_responsible_2023}, Future Ripples structures the generative process itself, enabling participants to trace these interconnected consequences in situ as they unfold during the workshop. This makes Future Ripples particularly suited to MST, an emerging domain where consequences remain highly uncertain and span technical, social, ethical, and legal dimensions that no single expert group can fully anticipate.}

\subsection{Calls for More Speculations on Mediated Social Touch}

\textcolor{black}{Existing speculative work on MST has begun to address this space from different angles: a digital social touch toolkit was proposed to explore attitudes towards the ethics of touch\cite{jung_digital_2024}; a literature review identified consent, trust, and control as three interconnected ethical concerns \cite{jewitt_2019_DigitalTouchEthics}; and a similar call advocates for emphasising social and experiential aspects of future MST interactions \cite{huisman_2022_interaction}. Most recently, a survey of 29 haptic experts mapped the psychological, technological, and societal opportunities and risks of MST \cite{barbareschi_moving_2025}. However, two gaps remain. First, existing empirical work on MST has primarily drawn on expert perspectives, a limitation that prior work has highlighted as risking a narrower scope of envisioned futures \cite{jewitt_2019_DigitalTouchEthics, jewitt_2021_manifesto, van_erp_editorial_2023}. Our work addresses this by using Future Ripples to bring together potential users, domain experts, and haptics researchers to build more holistically and socially grounded visions for MST. While Future Ripples has been applied to domains such as the metaverse \cite{hohendanner_metaverse_2024}, generative AI \cite{hohendanner_initiating_2025, holopainen_infinity_2025}, and more-than-human design \cite{rosen_peering_2024}, it has not yet been used to speculate about MST. Second, although these studies have identified broad ethical concerns, it remains unclear how haptics researchers might concretely act on them. We address this by adopting the RSR pipeline as an organising lens to trace where specific challenges manifest across the MST design process, helping researchers and designers to locate intervention points and consider how they might be addressed.}

\begin{figure*}[htb]
    \centering
    \includegraphics[width=1\linewidth]{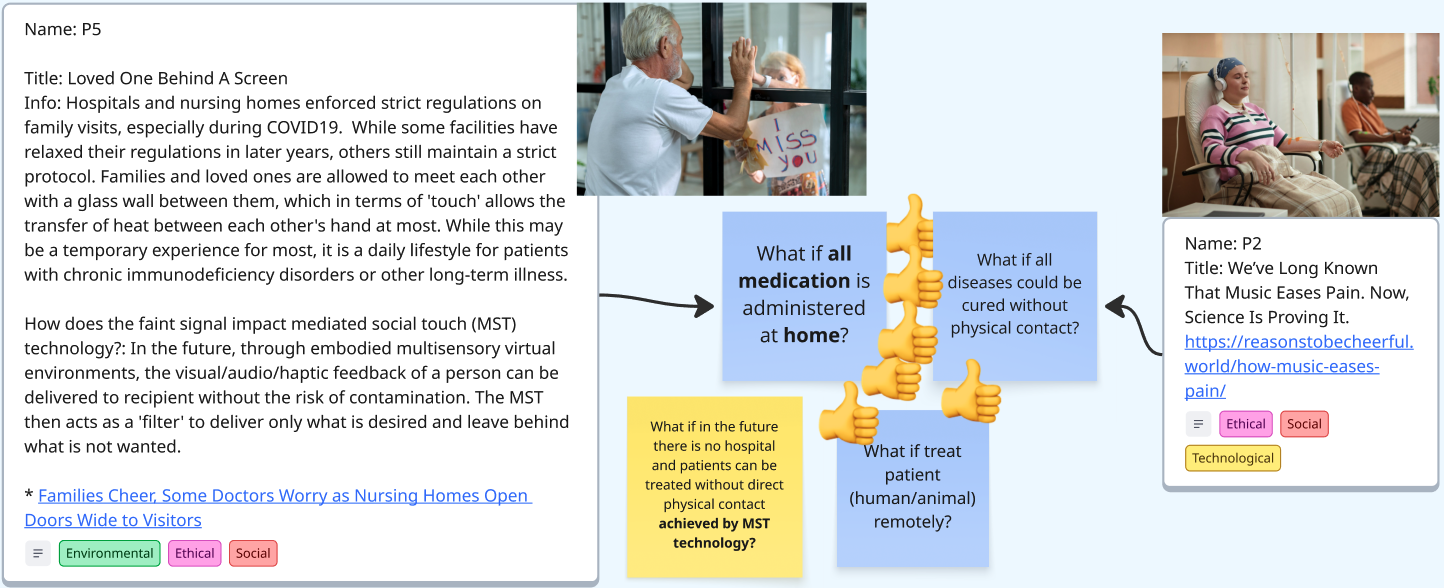}
    \caption{Example weak signals (left \& right, white) presented by participants that were used to inform the what-ifs (centre) in WS1. Blue post-its show the formulated what-ifs by participants. The yellow post-it shows the reformulated what-if with the help of the facilitators that is more closely linked to MST.}
    \Description{The figure shows two weak signals from participants, displayed on the left and right, each paired with photos and short descriptions. On the left, a scenario about family visits during COVID-19 highlights restrictions in nursing homes, illustrated with a photo of a person touching glass to reach a loved one. On the right, a scenario describes how music can ease pain, with a photo of patients in a clinic. Arrows point from these weak signals toward the centre, where post-it notes present what-if scenarios formulated by participants. Blue post-its pose questions such as “What if all diseases could be cured without physical contact?” and “What if treat patient (human/animal) remotely?” A yellow post-it refines this into a more MST-related scenario: “What if in the future there is no hospital and patients can be treated without direct physical contact achieved by MST technology?” At the top, a larger blue note asks, “What if all medication is administered at home?” The figure illustrates how participants’ weak signals were transformed into what-if questions that guided workshop exploration of mediated social touch.}
    \label{fig:WeakSignalExample}
\end{figure*}

\section{Method: Future Ripples Workshops}

To address our research questions, we ran a series of futuring workshops following the structure proposed by the \textit{Future Ripples} method~\cite{epp_reinventing_2022}. We chose this method for its ability to surface not only opportunities for technological innovations but also social, ethical, political, or environmental implications. As a speculation method orientated towards technology innovation and design, \textit{Future Ripples} aligns with our aim to critically engage with the impact of mediated social touch for human connection across distance. Each workshop session included a distinct participant group to provide perspectives from 
both potential users and technology designers on the futures of MST technologies. The workshops were structured iteratively, with insights from each workshop informing the next and supporting a cumulative and reflexive research process: Workshop 1 (WS1) gathered initial visions, expectations, concerns, and desires from everyday potential users of MST technology; Workshop 2 (WS2) drew on expertise from relevant professional fields to critically reflect on and expand the speculative futures and consequences generated in WS1; Workshop 3 (WS3) translated and built upon the cross-stakeholder insights generated in WS1 and WS2 with the expertise in technology and perception from haptic researchers. This structure allowed us to move from user-generated visions (WS1), through critical examination by domain experts (WS2), to design-oriented considerations with haptic researchers actively developing MST technologies (WS3) -- progressively grounding speculative futures in both societal concerns and technological feasibility.

\subsection{Workshop Activities}
At the beginning of each workshop, we provided a definition of ``haptic technology'', ``social touch'', and ``mediated social touch'' (MST) to ensure shared understanding, followed by a Q\&A session to clarify any questions. A \textit{Future Ripples} workshop typically consists of three main activities~\cite{epp_reinventing_2022}: Activity 1 (A1) grounds speculation in weak signals and trends,  \textcolor{black}{observable signs of change in the present that may point to possible future developments,} to formulate a ``what-if'' prompt; Activity 2 (A2) envisions direct and indirect consequences of the ``what-if'' through creating ``ripples''; and Activity 3 (A3) identifies strategies to pursue preferable or avoid undesirable futures. The ripples metaphor draws on a sea analogy: a pebble represents a ``what-if'' scenario, and once dropped into the sea, the ripples spreading outward symbolise its cascading consequences, capturing the uncertainty and volatility of possible futures. We adapted this structure by adding \textit{A0 Future Check-In} as an ice-breaking activity~\cite{werner_CoCreatingFuturesPractical_2025} and \textit{A3.2 Presentation \& Discussion} for participants to share their proposed strategies with the group. While A2 remained consistent across all workshops, we adjusted A1 and A3 to suit the distinct participant groups and goals of each session (see \autoref{tab:WorkshopActivities} for the full workshop structure and \autoref{fig:workshopComparsionTable} for the resulting what-if prompts). At the end of the workshops, we presented participants with a link to an online questionnaire asking about how they experienced the workshop, including personal challenges and learnings.

\subsubsection{Modifications to A1} \label{sec:ModificationsToA1}
In WS1, each participant brought two weak signals/trends from their personal and professional lives and formulated ``what-if'' prompts based on these signals. As a group, they voted on the most relevant prompt, which we then reformulated to link more closely to MST (see \autoref{fig:WeakSignalExample} for the progression from weak signals to the final ``what-if'' prompt). For WS2 and WS3, we shifted the focus from signal generation to critical examination. Rather than asking participants to bring new signals, we presented signals from WS1 spanning different MST applications: social interactions~\cite{wu_2025_HuggingSuit, teh_huggy_2008}, education/training~\cite{luo_adaptive_2024, shen_fluid_2023,antonyaPreservationCulturalHeritage2022a}, emotional support, and online collaboration. This allowed domain experts (WS2) and haptic researchers (WS3) to dedicate their time to critically reflecting on and expanding the user-generated futures (A2) and proposing actionable steps (A3), rather than starting from scratch. In WS3, held at an international haptics conference, we divided twelve participants into three groups, each focusing on a different application area from the signals above. We presented the respective trend to each group and collaboratively agreed on a ``what-if'' prompt, allowing haptic researchers' expertise to shape the starting point of speculation. This design enabled parallel exploration of multiple MST futures while grounding speculative visions in technological feasibility.

\subsubsection{Modifications to A3}
The original Future Ripples method suggests brainstorming actionable steps after plotting consequences on a likelihood/impact graph~\cite{epp_reinventing_2022}. In WS1, we followed this approach, with participants voting for the most impactful consequences and plotting them on this graph to identify those most crucial to address in the future. In WS2 and WS3, we replaced the open brainstorming with a stakeholder mapping template (inspired by \cite{hohendanner_initiating_2025}) to provide a more structured framework for identifying affected groups and possible strategies (see \autoref{fig:ActionableStepsTemplate} for details). This enabled domain experts and haptic researchers to draw on their professional knowledge in proposing actionable steps to maximise potential benefits and minimise threats.



\begin{figure*}[htb]
    \centering
    \includegraphics[width=1\linewidth]{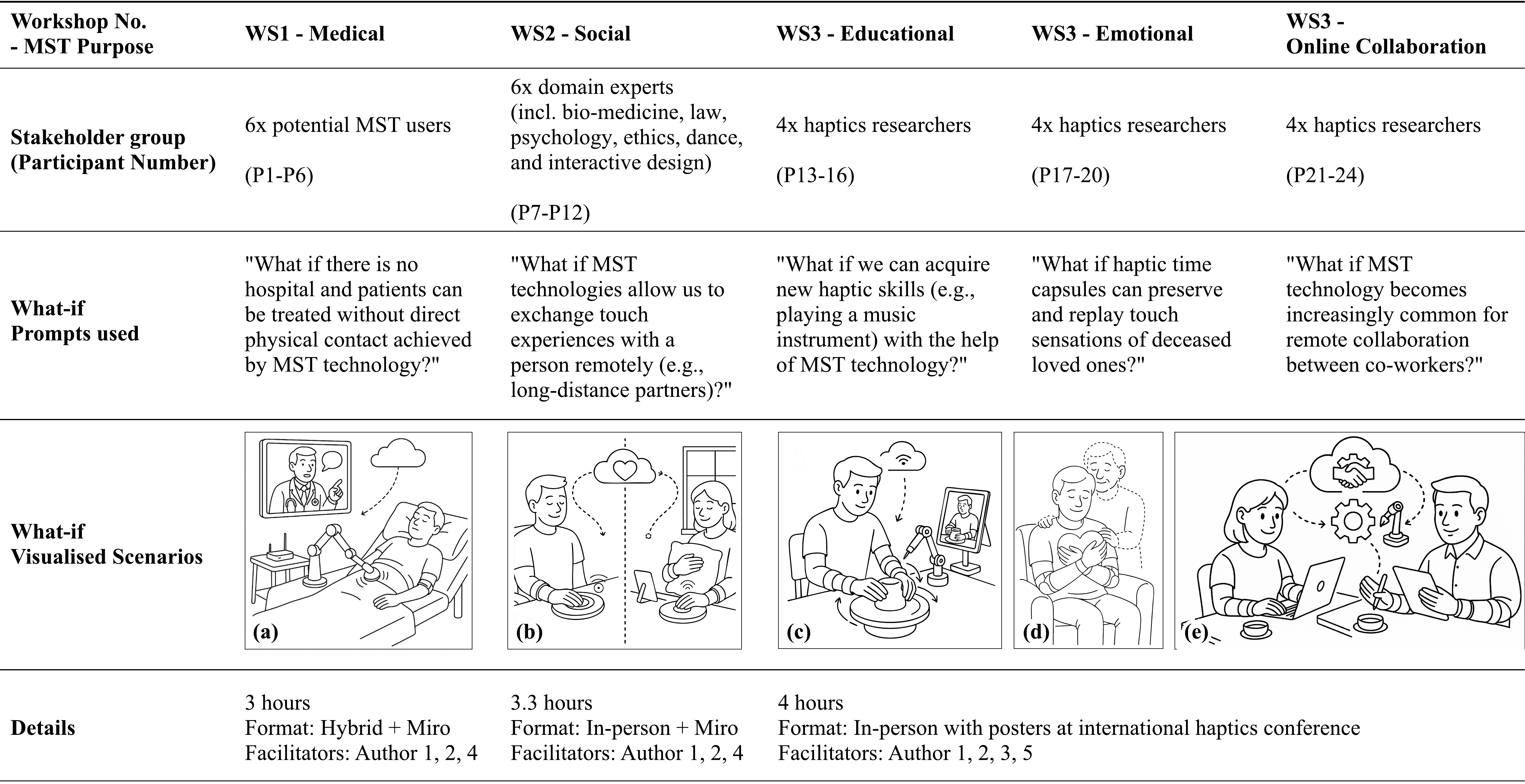}
    \caption{Summary of workshops. Based on the five ``what-if'' prompts formulated with participants in three workshops, we generated five \textbf{visualised scenarios} to help readers imagine and understand the different MST use contexts. Note: Images (a-e) generated using ChatGPT.}
    \label{fig:workshopComparsionTable}
    \Description{The figure summarises the three workshops conducted to explore mediated social touch (MST). Workshop WS1 focused on medical purposes, with six potential MST users (P1–P6) exploring the prompt: “What if there is no hospital and patients can be treated without direct physical contact achieved by MST technology?” Workshop WS2 addressed social purposes, involving six professionals from fields such as law, psychology, ethics, and interactive design (P7–P12), and used the prompt: “What if MST technologies allow us to exchange touch experiences with a person remotely (e.g., long-distance partners)?” Workshop WS3 explored educational, emotional, and collaboration purposes with 12 haptics researchers (P13–P24) and prompts such as acquiring haptic skills, replaying touch sensations of loved ones, and using MST for remote collaboration. All workshops included Futures check-in, presentation of weak signals, formulation of what-if prompts, creating future ripples, and selecting actionable steps, with some variation in format and presentation. WS1 and WS2 lasted 3.5 hours, while WS3 lasted 4 hours. WS1 was hybrid, WS2 and WS3 were in-person, and facilitation was provided by authors in different combinations.}
    
\end{figure*}

\subsection{Participants}

In total, we conducted three workshops (3, 3.3, and 4 hours) and speculated about five distinct purposes of future MST experiences with 24 participants (11 women, 13 men). WS1 was organised in hybrid form with most participants attending in-person and two remotely; WS2 was organised in-person with one of the facilitators joining remotely; and WS3 was organised completely in-person at a major, international conference on haptics research. Each workshop included a different group of stakeholders: six potential users in the first; six professionals from relevant fields (including bio-medicine, law, psychology, ethics, dance, and interactive design) in the second; and twelve haptics researchers in the third. \textcolor{black}{Participants ranged in age from 18 to 64, with the majority (11 of 24) in the 25--34 age group (see~\autoref{tab:participants} for a full breakdown). Among the twelve haptics researchers in WS3, four had more than 10 years of experience, one had 6--10 years, two had 4--6 years, and five had 1--3 years in haptics research. All had at least one year of experience with MST.} All participating haptics researchers had at least one year of experience with MST. \textcolor{black}{For WS1 and WS2, participants were recruited through workshop flyers and social media posts. For WS3, participants registered through the IEEE World Haptics Conference 2025 website\footnote{\url{https://2025.worldhaptics.org/Workshop-Tutorials/}}, where the session was advertised as part of the conference programme. At the beginning of each workshop, all participants were informed that the session was part of a research study and provided written consent before participating. This study received ethical approval from the University of Sydney Human Research Ethics Committee (HREC), protocol 2024/HE001380.} Participants took part in the workshops on a voluntary basis and were offered the opportunity to learn the Future Ripples method as an incentive. We summarise the three conducted workshops, including what-if prompts and participant groups in~\autoref{fig:workshopComparsionTable}. 

\begin{figure*}[htb]
    \centering
    \includegraphics[width=1\linewidth]{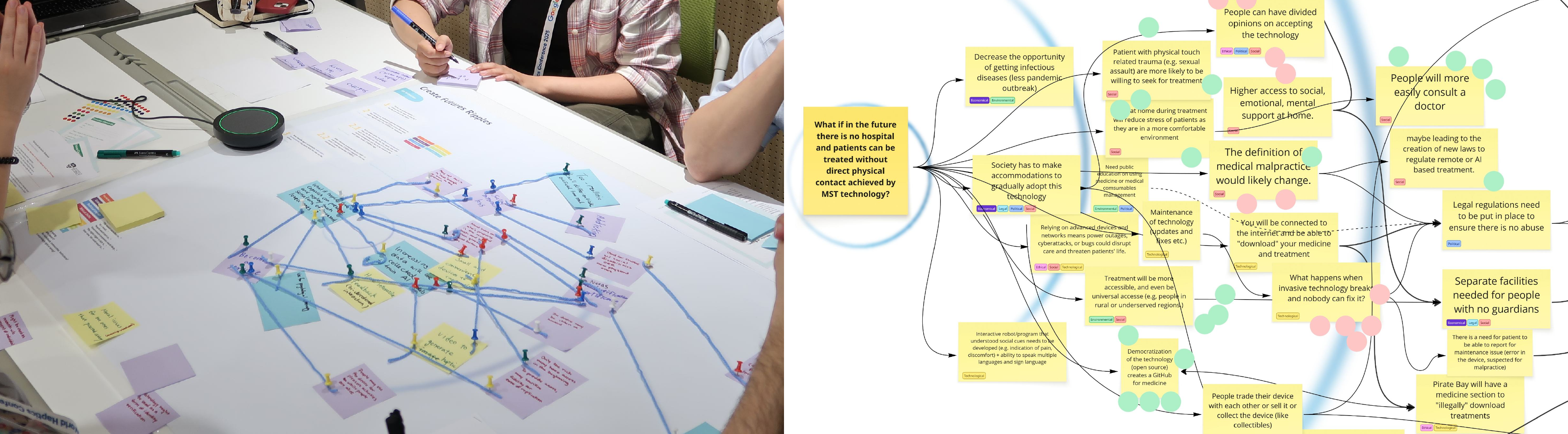}
    \caption{Physical creation of ripples during WS3 (left) and its digitised counterpart used in WS1 and WS2  (right).}    
    \Description{The figure contrasts two approaches to creating Future Ripples. On the left, a photo shows participants in WS3 physically placing colored post-its on a large printed template and connecting them with blue string secured by push pins, forming a network of consequences across the board. Hands, pens, and other workshop materials are visible around the table. On the right, a screenshot from Miro displays the digital version used in WS1 and WS2, where post-its are represented as yellow notes connected by curved lines into clusters. The side-by-side comparison illustrates how WS3 generated tangible artefacts that were later digitized to remain consistent with the online artefacts of WS1 and WS2.}
    \label{fig:AnalogDigitalRipples}
\end{figure*}

\subsection{Data Collection}

\paragraph{Video Footage \& Audio Recordings} 
All participants consented to the recording of video \& audio during the duration of the workshop to be used for data analysis purposes only. For WS1 and WS2, the video and audio of all participants were recorded through the Zoom meeting recording feature. In WS3, for each group, a laptop was set up at one end of the table so that all participants were within the built-in webcam's field of view. A USB speakerphone connected to each laptop captured the audio of participant discussions (see~\autoref{fig:AnalogDigitalRipples}, left).

\paragraph{Participant Output}
In WS1 and WS2, participants used their own laptops to access the Miro board with templates that facilitated all activities. For WS3, participant-filled post-its on a physical Future Ripples board (A2) and Actionable Steps board (A3) were collected and then digitised for analysis
(see~\autoref{fig:AnalogDigitalRipples}, right). 

\paragraph{Post-Workshop Questionnaire}
\textcolor{black}{At the end of each workshop, participants were invited to complete an online questionnaire with open-ended questions about their workshop experience, including key takeaways and which parts they found most enjoyable or challenging. Responses were used to complement the facilitators' reflections in Section \ref{methodsreflection}.}

\subsection{Data Analysis}

We conducted a thematic analysis, \textcolor{black}{ chosen for its flexibility in identifying patterns across different participant perspectives and its suitability for exploratory, qualitative data \cite{clarke_thematic_2014}. The ripples, composed of post-its written by participants during the workshops, present a visual mapping of interconnected consequences and served as the primary unit of coding (see~\autoref{fig:AnalogDigitalRipples}). During the workshops, participants first wrote their speculated consequences on post-its and then presented them verbally to the group. The post-its therefore represent the core concepts participants chose to foreground. The transcripts provided richer contextual detail and were used to supplement the coding and to source participant quotes that illustrate the identified themes.}

We coded the ripples in two cycles based on \cite{wicks_coding_2017}. In the first cycle, we applied descriptive coding, summarising the speculated consequence with a short word or phrase. In a second cycle, we applied focused coding to identify patterns and recurring themes among the descriptive codes. Authors 1 and 2 coded all five ripples, while authors 3 and 5 joined the coding process for the ripples from the third workshop in which they also served as facilitators. \textcolor{black}{After individual coding, we met to compare and discuss codes, iteratively grouping related codes into candidate themes. These candidate themes were then refined through multiple meetings with all authors, where we reviewed whether each theme was supported by sufficient data across the ripples and transcripts and whether the themes collectively captured the range of participants' speculations.}

In Section \ref{findings}, we present the opportunities and threats \textcolor{black}{that emerged across the three workshops, synthesised into themes that capture common patterns while preserving the range of viewpoints across stakeholder groups. This responds to our first research question on how various stakeholders imagine the potential futures of MST.
Section \ref{discussion} then addresses our second research question by identifying challenges distinctive to MST and adopting the RSR pipeline as an analytic lens to consider how they might be addressed.}

For the methodological reflections, authors 1, 2 and 3 first reflected individually about their experiences organising and conducting the workshops and then met to share these among themselves while considering the workshop participants' feedback collected from the online questionnaire. A summary is presented in Section \ref{methodsreflection}.

\section{Findings: Anticipated Opportunities and Threats}
\label{findings}

Throughout the workshops, participants envisioned MST technologies as simultaneously enabling new opportunities and posing significant threats. Their speculations were sometimes anchored in the current technology landscape, with recurring references to social platforms and AI. Accordingly, many anticipated that the emergence of MST not only brings new applications but also exacerbates issues caused by existing technologies, such as bullying and dependence.

We also observed \textcolor{black}{that participants frequently imagined MST beyond remote haptic feedback alone, at times extending their speculations to include mediated movement or remote actuation.} This was particularly apparent in accounts framing MST as a `superpower' or speculations of scenarios and consequences that would include touch as a key, though not exclusive, modality. \textcolor{black}{We retained these broader speculations in our analysis, as touch remained central to all envisioned scenarios, and narrowing the scope to a strict definition of MST would have excluded insights that participants considered integral to the future of MST.}

Our analysis resulted in the identification of the following four themes, reflecting some of these associations: (1) Digitised Touch Catalyses Emergent Services, (2) Haptic Big Data Enables Data-Driven MST, (3) Remote Touch Complicates Consent and Exposes Users to Physical Harassment, and (4) Asynchronous Touch Reshapes Social Touch Practices. \textcolor{black}{Themes 1 and 2 are presented with explicit reference to what-if scenarios, as the envisioned impacts were closely tied to specific application contexts. Themes 3 and 4, by contrast, reflect concerns that emerged consistently across multiple workshops and contexts, suggesting these are broader challenges inherent to MST and not tied to any single application domain. Rather than organising findings by context, we chose to organise by theme to highlight these cross-cutting patterns.}

In the following sections, we present these themes in detail, accompanied by the scenarios explored across the three workshops. As quotes are often scenario-specific, and to further improve contextualisation of findings, we annotate quotations as ``W[i], MST Purpose (see~\autoref{fig:workshopComparsionTable}, i)'' to improve contextualisation. Readers can refer to~\autoref{fig:workshopComparsionTable} for both the visualised scenarios and the what-if prompts that informed each finding. For example, "WS1, Medical" refers to Workshop 1, where MST was envisioned to be applied in a medical scenario. \textcolor{black}{Where relevant, we reference prior work to contextualise participants' speculations within existing research, though the findings reported here are grounded in workshop data.}


\subsection{Digitised Touch Catalyses Emergent Services}

By turning touch into a recordable, transferrable, and reusable resource, digitised touch catalyses the emergence of novel service models across healthcare, education, and beyond. Workshop participants imagined MST as enabling entirely new forms of services while also recognising its disruptive impact on existing professions. These possibilities were articulated through speculative service scenarios (what-ifs, listed under~\autoref{fig:workshopComparsionTable}) that simultaneously emphasised expanded access and disruptions to existing professional structures. For example, future MST was envisioned as making touch-based therapy and training more widely available, from remote exposure therapy for trauma or autism to democratised skill acquisition.

In WS1, Medical (see~\autoref{fig:workshopComparsionTable}, a), participants envisioned how MST could transform healthcare services by enabling care to take place beyond traditional hospital settings. The touch of a healthcare provider could be captured by sensor-equipped gloves, encoded as data, and reproduced on the patient’s side with haptic devices such as robotic arms or smart beds. \textcolor{black}{Similar configurations have been explored in prior work on exoskeleton-mediated physical interaction~\cite{Vianello_Exoskeleton_2024}.} In this way, a massage, palpation, or even the comfort of hand-holding could be transmitted across distance. Once digitised, ``MST treatments'' could also be stored and shared. Patients could ``download'' therapeutic routines, and communities could build a ``GitHub of medicine'', remixing and redistributing protocols: ``\textit{Maybe it's good to think about it in an open-source way [...] people can also work on this open-source to make it even better. So, it's kind of like community supported}''~(P3).

In WS3, Educational (see~\autoref{fig:workshopComparsionTable}, c), participants envisioned how MST could transform skill training. \textcolor{black}{Complex touch-based practices, such as violin instruction or surgical techniques, could be recorded and replayed, making them more accessible and affordable beyond traditional co-located instruction. Recent haptic systems are beginning to enable such possibilities, from gloves that let users feel individual strings of a virtual violin to improve their technique \cite{shen_fluid_2023}, to smart gloves that transfer piano-playing skills from expert to novice through tactile guidance \cite{luo_adaptive_2024}.} ``\textit{Suddenly many people can learn violin [...] it may become easier for many people to have access to previously niche skills. I guess, like, if you have one-to-one violin lessons, then it costs a lot of money, but suddenly many people can learn the violin}'' (P15). Others highlighted applications in rare medical training: ``\textit{Maybe the training of [treating] rare diseases might become possible}'' (P14). Yet such digitisation could undermine traditional apprenticeship and professional cultures. As one participant put it: “\textit{The main surgeon was like a god [...] now with endoscopic cameras, everyone can see. Maybe the same thing happens with MST}” (P13).

Building on this potential, participants shared a somewhat utopian vision in which MST reshapes how touch-related skills are produced and shared. As one noted, ``\textit{People can study [...] what they want or become good at what they want. Not only with [what] you have access to...}'' (P15). Participants further imagined this to be implemented through online databases that store and distribute professional movements—``\textit{a website that stored all the professional movements, and they can download it}'' (P16)—echoing the previously described ‘GitHub of medicine’ concept. At the level of governance, this increase in access and availability of the traditionally 'invisible' touch-related skills via MST may consequently increase the transparency of specialised practices. Thus the authority of professionals traditionally associated with their know-how accumulated over many years of experience may relatively decrease as a result of transparency: "\textit{Professionals lose authority [that] comes after transparency}" (P15).

Conversely, by reconfiguring how training is delivered and accessed as a service, MST could cause communities built around certain skill training to diminish: ``\textit{But then you will not have these communities anymore. It would just be you learning from, like, the social touch technology.}'' (P15), "\textit{They may lose the traditional way to learn or teach the Shokunin (craftsman). [...] The culture of the Shokunin may be gone...}" (P16), and ``\textit{[...] because of the pure intermediate [...] social touch, the face-to-face communication is not truly necessary for the training. So some kinds of culture may disappear}'' (P14). Additionally, participants expressed uncertainty about the impact of the newly envisioned teaching culture as a consequence of MST. Longitudinal comparisons between traditional and MST-based teaching approaches would be necessary to anticipate the long-term opportunities and threats of each approach, as mentioned by a participant: ``\textit{I don't think it's been studied so much how that kind of [MST] training is better [...] If I want to start doing clay, I can just start immediately with putting it on the wheel. But maybe it's actually not good to start like that. [...] It's not sure if it's like positive or negative [impacts], we don't know yet}'' (P15).


Taken together, these emerging services imply shifting roles and responsibilities. MST was seen as simultaneously enabling empowerment, access, and creativity through new services, while also introducing challenges of dependency and exclusion. 

\subsection{Haptic Big Data Enables Data-Driven MST}
Similar to familiar technologies involving interfaces for inter-user communication (e.g., smartphones, smartwatches, VR headsets, etc.), MST usage logs from a population of users can accumulate over time to form big data. Participants naturally envisioned the emergence of AI models trained on these data to become capable of identifying user intent, personal preferences, predicted responses, etc. However, due to the personal nature of touch data, participants raised concerns about data ownership, privacy, and authenticity resulting from the potential integration of AI with MST, perceiving it as both an enabler and disruptor.

In WS3, Educational (see~\autoref{fig:workshopComparsionTable}, c), MST was envisioned to serve as training data accelerating AI in learning human skills, leading to expedited replacement of human trainers: ``\textit{The music trainer can be easily replaced by AI because MST means that the training skills can be digitised, meaning that [they] can be easily learned by AIs}'' (P14). On the other hand, based on learned patterns of socially acceptable communication between MST users, AI was envisioned to reduce harassment via MST by serving as an intermediary layer during interaction to ``\textit{control the boundaries}'' (P13 and P24) between acceptable and harassing MST: ``\textit{teachers → AI → [MST devices] → learners}'' (P14).

In WS3, Emotional (see~\autoref{fig:workshopComparsionTable}, d), participants speculated about recording and replaying hugs from deceased loved ones. When such intimate haptic data is stored and potentially used to train an AI model, questions of data ownership become particularly fraught: ``\textit{Whose property is that hug? Do we need laws to define it?}'' (P18). \textcolor{black}{Beyond questions of individual data ownership, participants raised concerns about consent over how data is transformed.} They noted that AI could generate artificial touch even without direct haptic input, drawing instead from other data sources such as videos or images: ``\textit{He [Michael Jackson] is not there anymore, but he has many videos, many images}'' (P20). This possibility raises additional concerns, as individuals may never have consented to their data being transformed into haptic experiences. \textcolor{black}{At a broader scale,} the prospect of haptic data continuously feeding AI models was thus perceived as reinforcing surveillance economies: ``\textit{Extensive data collection will be needed [...] this is feeding predictive AI and new forms of surveillance}'' (P24). The concern is that ongoing data collection could enable corporations to monitor users and target them based on inferred intentions or behaviours. Given the intimate nature of touch, such data could expose users to particularly invasive forms of profiling, such as revealing emotional states, health conditions, or the nature of their personal relationships.

In sum, haptic big data was seen as the foundation enabling AI integration, amplifying MST’s potential, expanding access, safeguarding users, and even generating artificial touch. Yet this integration also raises profound ethical questions about authenticity, surveillance, and the displacement of human expertise. Among these, consent emerges as a particularly complex challenge, which becomes even more pressing when touch can be transmitted remotely.

\subsection{Remote Touch Complicates Consent and Exposes Users to Physical Harassment}
As a major concern, participants expressed that MST could be used against users' consent and be misused for physical harassment and violence. As one participant noted, ``\textit{People can use haptic devices to send inappropriate messages, harassment, [and] bullying}'' (P23), with others warning of consequences such as ``\textit{increased bullying at work}'' (P22) and ``[mental] illness [caused by bullying] increase'' (P24). \textcolor{black}{Just as users can send unwanted text or image messages on existing platforms, MST would enable the delivery of unwanted tactile sensations, extending familiar forms of digital harassment into the physical domain.} Beyond "MST bullying," participants also speculated about more extreme misuse. One participant described that ``\textit{MST allows us to enact violence, punishment or torture}'' (P11), while another warned that such technology could enable assaults ``at scale'' (P10) and be hard to regulate. To address these concerns, participants highlighted that consent and agency are critical considerations for MST design. As P7 put it, ``\textit{MST can cause problems with establishing consent},'' and P15 added, ``\textit{It has to do with agency, like how much control [users] feel over the situation.}'' Participants emphasised the importance of user autonomy, including the right to disengage: ``\textit{I also think about the right [to] turn it off […] it’s like an extra [protection mechanism] for mental health}'' (P23). These concerns underscore that remote touch creates new avenues for consent violations, as harassment occurs precisely when touch is delivered without the recipient's agreement.

Beyond the risk of intentional misuse, participants discussed how consent in MST is inherently ambiguous due to limited visual cues available in remote interactions. In face-to-face settings, certain touches carry implicit consent: as one participant explained, ``\textit{Touch is not always about consent […] if you’re patting a friend on the back because they did something good, you don’t ask for consent}'' (P12). Yet when mediated through technology, these blurred boundaries become harder to manage and adjudicate. As P10 noted, ``\textit{The jury can be asked to consider any active steps that the defendant took to find out [whether the other person was consenting]}''. In MST the 'touch giver' has fewer visual or contextual signals, making it harder to recognise the consent from another. Such concerns highlight the unique challenge of negotiating consent when touch occurs remotely.

\begin{figure*}[htbp]
    \centering
    \includegraphics[width=\linewidth]{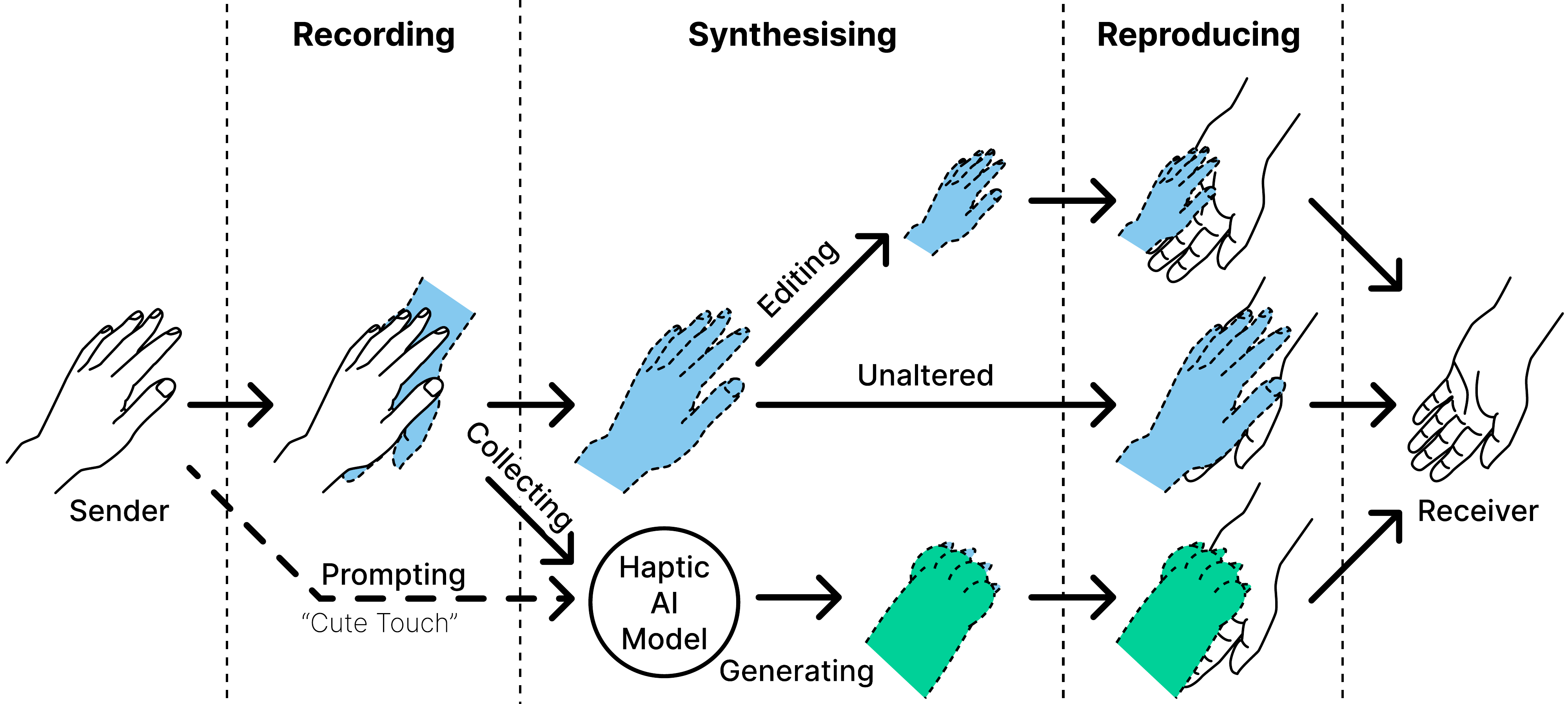}
    \caption{Visualisation of RSR Pipeline. Building on a previous paper~\cite{Taguchi_2023_Multichannel}, MST technology includes three phases: recording, synthesising, and reproducing (RSR). Recording means to capture the natural haptic sensation that happens in the real world; synthesising means to edit the recorded haptics or generate artificial ones (such as the sensation of an animal skin); reproducing means to replay recorded or synthesised haptic sensation through MST interfaces.} 
    \Description{The figure illustrates the Recording–Synthesising–Reproducing (RSR) pipeline for mediated social touch (MST). On the left, a sender’s hand performs a touch that is recorded with a blue overlay. This recorded touch can be collected and passed to a haptic AI model, which can either reproduce it unaltered, generate a new form represented as a green cat’s paw, or allow the touch to be edited—in this case, shown reduced in scale. In the reproducing stage on the right, the receiver’s hand experiences three types of MST reproduction: edited, unaltered, and generated. A dashed line labeled “Prompting” at the start also indicates that touch may be initiated by instruction, such as a “Cute Touch.” The figure demonstrates how MST can record real touches, manipulate them, or generate synthetic versions before transmitting them to a receiver.}
    \label{fig:RSRPipeline}
\end{figure*}

\subsection{Asynchronous Touch Reshapes Social Touch Practices}
Social touch typically unfolds in real time. MST can potentially introduce the opportunities for asynchronous touch: recording, storing, and replaying social touch across time. Participants anticipated that this shift would reshape social touch practices, gradually pulling people away from real-time, co-present interaction. One participant drew an analogy to video media: ``\textit{I think the social interaction […] would be like [the video] on YouTube. You could save and watch it later. So maybe you wouldn't really live in the moment.}'' (P19). As this analogy suggests, when touch can be deferred, the immediacy that once defined social touch may be lost.

However, this possibility could lead people to substitute recorded social touch for real-time interaction. Participants imagined scenarios where people would prefer sending pre-recorded touch rather than engaging in the moment. As one participant put it: ``\textit{They're like, I really don't want that much [social] interaction. Did you just play what I gave you? Can I just send you the [MST] file I recorded last time, please?}'' (P18). Here, the participant imagines a future where pre-recorded social touch becomes a means of avoiding the demands of real-time engagement.

When real-time interaction is no longer required, physical co-presence may also become unnecessary. Participants further envisioned that such substitution could reduce co-present interaction more broadly. P19 illustrated a dystopian scenario: ``\textit{It's kind of like a cyberpunk vibe. People [are] getting addicted to this technology, so [it's] kind of related to less social interaction. Maybe they just stay at home and constantly use this device}''. In this vision, asynchronous touch leads not merely to convenience but to withdrawal from embodied social life.

Over time, this reduced exposure to direct social touch could lead people to feel estranged from it. Referring to behavioural changes observed after the COVID-19 pandemic, participants noted that some people felt foreign or even aversive to real human touch after prolonged physical distancing (P7, P8). Participants anticipated a similar effect from habitual MST use: ``\textit{people have less desire to touch each other}'' (P23). Ultimately, this estrangement was seen as carrying deeper psychological consequences. As one participant reflected: ``\textit{I think in the beginning people will just feel excited about having a way of touching […] But at the end I think it could lead people to loneliness, depression, and anxiety. […] I don't think there is anything that can actually replace human touch}'' (P8).

\section{Discussion: The Distinctive Challenges of Mediated Social Touch}
\label{discussion}


In the previous section, we addressed RQ1 by outlining the opportunities and threats that participants envisioned across five MST application contexts. Many of these concerns echo threats associated with other emerging technologies~\cite{Puntoni_DataMiuse_2021,Roberts_SocialIsolation_2025}. However, participants also pointed to challenges that are distinctive to MST, arising from the uniquely embodied, intimate, and ephemeral nature of social touch. To address RQ2, we extract three such challenges and their corresponding design implications: (1) Encoding Touch-Based Expertise without Eroding Apprenticeship, \textcolor{black}{which emerged primarily from the medical (WS1) and educational (WS3) scenarios;} (2) Establishing Appropriate Consent Negotiation Protocols, \textcolor{black}{which was raised across the social (WS2), educational (WS3), and online collaboration (WS3) contexts;} and (3) Moderating the Persistence of Touch, \textcolor{black}{which surfaced most prominently in the emotional scenario (WS3) but was also echoed in the social (WS2) and online collaboration (WS3) contexts.} Each of them represents a fundamental transformation that occurs when MST is adopted widely.

To mitigate these challenges, we adopt the RSR (recording, synthesising, reproducing) pipeline as an analytical lens~\cite{erk_2015_effects,minamizawa_techtile_2012,Taguchi_2023_Multichannel} (see \autoref{fig:RSRPipeline}), informed by the actionable steps that participants proposed in Activity 3 (A3) of WS2 and WS3. We broaden this pipeline beyond its original focus on ``editing and replaying'' vibrotactile content to encompass haptic data that can be transformed, combined, or artificially generated, such as training material for generative AI models. In our usage, ``synthesising'' includes both editing recorded sensations and generating new ones, while ``reproducing'' refers to delivering haptic sensations through MST interfaces. This mapping serves to localise responsibility and identify intervention points where haptics researchers and designers can implement mitigation strategies. Each subsection thus examines both what makes a challenge distinctive and how it might be addressed across the RSR pipeline.

\subsection{Challenge of Encoding Touch-Based Expertise without Eroding Apprenticeship}

Our findings revealed a tension about the possible change of touch-based skill transfer via MST. On one hand, participants envisioned digitised touch democratising access to professional skills. Skills that typically require prolonged, in-person apprenticeships, such as surgery or traditional crafts, might become learnable by more people without being physically co-present with an expert. On the other hand, they anticipated that encoding such knowledge might displace the relational structures through which it is developed and transmitted: apprenticeship cultures eroded, learning communities diminished, and skill acquisition reduced to solitary interaction with technology. This tension between democratised access and relational erosion raises a question that existing literature has yet to fully address.

As knowledge scholars have long noted, "we know more than we can tell": the craftsman's sense of "right" pressure, the surgeon's feel for tissue, or the musician's shaping of a phrase cannot be fully articulated in explicit words~\cite{Polanyi_TacitDimension_1997}. This kind of embodied knowledge is built up through practice and was traditionally passed down only through co-present apprenticeship in communities of practice~\cite{Lave_LegitimatePeripheralParticipation_1991}. Recent research underscores MST's capacity to facilitate the transfer of manual skills across various domains, ranging from crafting to medicine~\cite{barbareschi_moving_2025}; yet this framing foregrounds the democratisation of access without examining how encoding might alter the tacit and relational dimensions of expertise itself. The design challenge is not only to accurately capture tactile signals but also to encode touch-based knowledge without disconnecting it from the communities of practice in which it is taught and learned.

\noindent{\textbf{Design Implication Across RSR: }} The RSR pipeline offers a lens for identifying where this challenge manifests and how it might be addressed. The most direct concern emerges at the \textbf{reproducing} stage: how can encoded touch be delivered without rendering learning a de-contextualised experience? Our participants voiced precisely this worry—that skill acquisition might become "\textit{just you learning from the social touch technology}" (P15), severed from the human relationships that have traditionally sustained it. Designing MST applications for such educational contexts should consider hybrid, multimodal learning experiences, in which recorded touch is integrated with multisensory interactions. Such designs might also support shared haptic learning spaces that connect learners around common tactile experiences, helping to sustain a sense of community even when the co-present teacher is absent.

The \textbf{recording and synthesising} stages present less immediate but equally important considerations. What relational and contextual information should be captured alongside the tactile signal itself—the teacher's identity, the pedagogical context, the intended recipient? And how might editorial or algorithmic processing preserve rather than strip away the markers of human relationship embedded in the original touch? These remain open questions for haptics researchers, but attending to them at each stage of the pipeline may help ensure that digitised touch-based knowledge retains its connection to the communities through which it was constituted.

\begin{figure*}[htbp]
    \centering
    \includegraphics[width=1\linewidth]{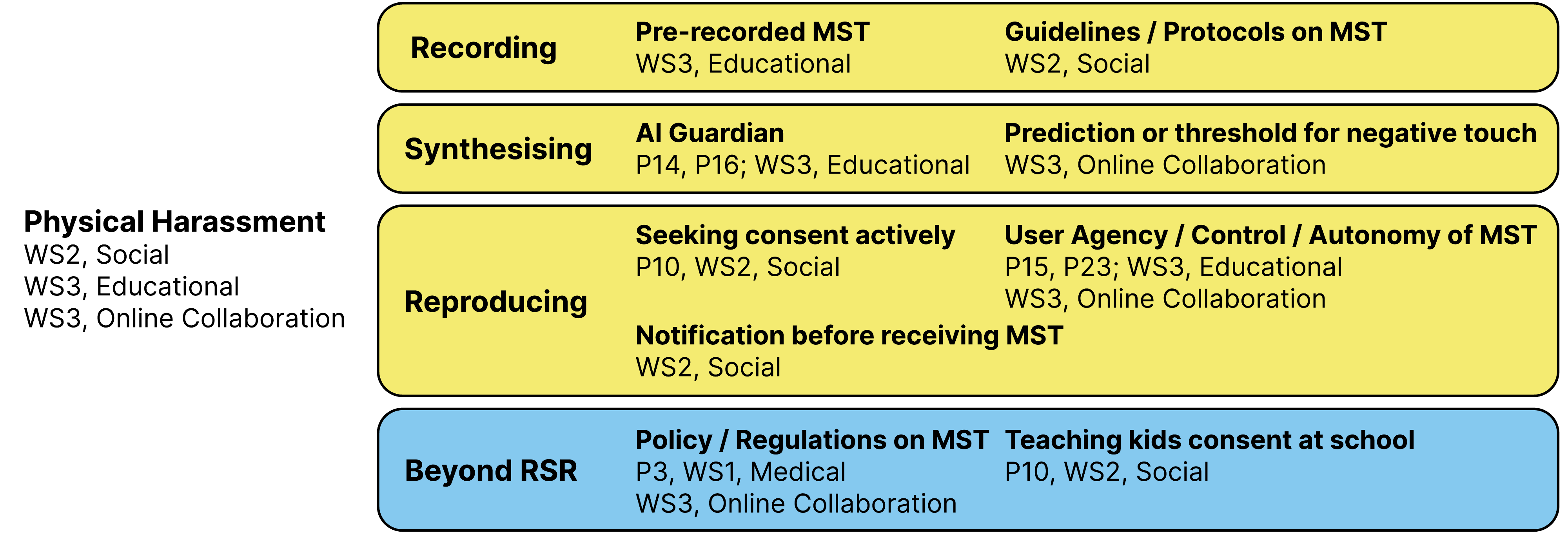}
    \caption{Mapping participants' actionable steps to different stages of the RSR pipeline. From top to bottom, physical harassment on the receiver and privacy intrusion on the sender of MST.}
    \Description{The figure presents participants’ proposed responses to an ethical challenge in mediated social touch (MST): physical harassment. For harassment, responses are distributed across the RSR pipeline. In the recording stage, participants suggested pre-recorded MST and guidelines or protocols. In the synthesising stage, they proposed an AI Guardian and thresholds to detect negative touch. In the reproducing stage, actions included seeking consent, maintaining user agency and control, and notifying the receiver before touch occurs. Beyond RSR, participants recommended policies and regulations and teaching consent at school.}
    \label{fig:ResponsesToChallenges}
\end{figure*}

\subsection{Challenge of Establishing Appropriate Consent Negotiation Protocols}
Sharing across domain experts and haptics researchers, participants were deeply concerned about MST being misused for harassment and unwanted touch. They emphasised the need for consent mechanisms that give users control over their MST experience. Yet they also recognised that consent in social touch is not always explicit—certain touches, such as patting a friend on the back, typically occur without verbal agreement. This creates a tension between protecting users from unwanted touch and preserving the spontaneity that characterises everyday social touch.

As previous research highlighted, issues of consent that are already complex in direct social touch become amplified when touch does not have to be synchronous or reciprocal~\cite{jewitt_2019_DigitalTouchEthics}. Research has shown that social touch is a temporally organised practice, coordinated through embodied initiations and responses—gaze, spatial proximity, body posture, and real-time adjustment~\cite{Cekaite_InteractionalApproachTouch_2020}. Through these contextual cues, the touching agent is continuously guided by these contextual cues to gauge whether social touch is welcome~\cite{ley_2021_touchingAtDistance}. MST disrupts this process by removing these cues. The design challenge, then, is how to design consent mechanisms for MST that can substitute for the contextual cues through which consent is negotiated in face-to-face interaction, without requiring users to approve every touch before it happens—which would undermine the spontaneity that characterises social touch.

\noindent{\textbf{Design Implications Across RSR: }}
Physical harassment was frequently discussed throughout WS2 and WS3 during Activity 3, providing rich data on how participants envisioned addressing unwanted touch through MST systems. We therefore map their proposed actionable steps onto the RSR pipeline (see~\autoref{fig:ResponsesToChallenges}). At the \textbf{recording} stage, participants proposed that pre-recorded MST could reduce harassment risk by introducing a time gap between sender and receiver, allowing platforms to moderate haptic content before delivery. Guidelines and protocols specifying what kinds of touch are permissible to record could also be established at this stage. At the \textbf{synthesising} stage, participants envisioned an "AI guardian" (P14, P16) trained to detect and filter harmful touch patterns before they reach recipients. At the \textbf{reproducing} stage, the focus shifts to recipient control: notification systems, the ability to accept or decline incoming MST, and preferences about who can send touch. As~\citet{cornelio_responsible_2023} argued, "the user must have full control of any tactile experience they receive."

\textcolor{black}{However, these moderation strategies face significant practical challenges. Current content moderation on social media platforms has proven difficult to implement effectively, even for text and images where automated detection is more mature \cite{Singhal_Sok_2023, Zhang_Debate_2024}. MST presents additional complexity: as our participants noted, the same touch can carry different meanings depending on the relationship between sender and receiver (P12), making it difficult to establish universal thresholds for what constitutes harmful touch. This suggests that purely automated approaches such as the "AI guardian" may require access to relational and contextual information that is difficult to capture computationally.}

Beyond the RSR pipeline, participants recognised that technical solutions alone are insufficient. Policy frameworks and regulations were considered essential to establish legal standards for MST consent, while education about MST consent was seen as equally important (P10). These complementary efforts highlight that addressing consent in MST requires collaboration across technical, legal, and educational dimensions. \textcolor{black}{Yet these strategies primarily address protecting users from unwanted touch. How to implement such safeguards without undermining the spontaneity of social touch remains underexplored.}

\subsection{Challenge of Moderating the Persistence of Touch}
Participants envisioned futures where touch could be recorded, stored, and replayed across time, much like video on YouTube. While this persistence offers opportunities such as preserving the touch of deceased loved ones, participants also expressed concerns that people might substitute MST touch for real-time interaction (see what-if in~\autoref{fig:workshopComparsionTable}, d). Over time, this could lead to estrangement from co-present social touch. This creates a tension between the benefits of preserving touch across time and the risk of diminishing its immediacy and relational presence.

Social touch is inherently temporal, with meaning emerging in the moment-by-moment unfolding of interaction and disappearing once the contact ends~\cite{Cekaite_InteractionalApproachTouch_2020}. When social touch can be recorded and replayed, it shifts from a shared, mutual exchange to an individual experience~\cite{jewitt_2019_DigitalTouchEthics}. This positions persistent touch between two possibilities: a meaningful way to preserve connection across time or an isolating substitute for embodied interaction. The design challenge, then, is how to leverage the benefits of persistent touch without eroding the temporal and relational qualities that give social touch its meaning.

\noindent{\textbf{Design Implications Across RSR: }}
The RSR pipeline helps identify where this challenge manifests. At the \textbf{recording} stage, the very act of capturing touch transforms it from a fleeting moment into a persistent artefact—raising questions about whether touch should be recorded by default or only with explicit intent. At the \textbf{synthesising} stage, recorded touch can be edited, combined, or generated artificially. When a hug can be algorithmically enhanced or generated from video data of someone who never consented to such use, questions of ownership of MST arise. At the \textbf{reproducing} stage, the ability to replay touch on demand may shift users' relationship with real-time interaction. Participants imagined scenarios where people would prefer to send a pre-recorded file rather than engaging in the moment.

Unlike physical harassment, where participants proposed concrete safeguards at each pipeline stage, the challenge of persistence received fewer actionable responses. This may reflect the novelty of the concern—touch has never before been recordable, so frameworks for managing persistent touch remain underdeveloped. Yet, the implications are significant: as one participant reflected, habitual MST use could ultimately lead to ``\textit{loneliness, depression, and anxiety}'' (P8) by substituting mediated social touch for embodied connection. Whether persistent touch will complement or compete with real-time interaction remains an open question—one that requires longitudinal research to understand how recording and replaying touch shapes social practices over time. Designers must consider not only whether to enable persistence but also how to frame it as a supplement to embodied touch rather than a replacement.

\section{Speculating About Touch: Methodological Reflections}
\label{methodsreflection}

Drawing on our experiences as facilitators of three Future Ripples workshops, we reflect on how methodological decisions may shape the ways participants imagine futures of mediated social touch. Rather than treating the method as neutral, our reflections highlight how facilitation choices influence which futures become imaginable, discussable, and actionable. 

One central tension concerned how broadly or narrowly the initial what-if prompt was framed. Broader prompts tend to encourage divergent and future thinking, allowing participants to explore more diverse scenarios. At the same time, they can dilute focus and lead discussions away from the specific concerns that motivate workshop goals. More narrowly framed prompts provide clearer direction and help participants stay aligned with workshop aims, yet they may also limit speculative range and constrain creativity. This tension became particularly visible in WS1, where the initial what-if was revised during the workshop to link it closer to MST. We started with the framing ``\textit{What if in the future there is no hospital and patients can be treated without direct physical contact}?'' but quickly ended up in too wide imaginations, and the focus of MST got lost. Hence, we rephrased in the second round of brainstorming by linking it directly to MST technology, as shown on the yellow post-it in \autoref{fig:WeakSignalExample}. Such moments revealed that the starting question functions as more than a practical entry point. It influences what kinds of futures feel relevant to explore by participants. While the STEEPLE categories are meant to help participants broaden their imagination even if the starting point is more concrete ("Think of social, technological, environmental, ethical, political, legal, or economic implications"), it is still a crucial step to balance openness and direction of the starting prompt. We found it useful to scaffold broader prompts with examples. When we asked participants to formulate their own what-if prompt, we gave some pointers: \textit{How does it provoke or challenge existing, unspoken assumptions? Does it lead to interesting questions of 'then what'? Is it constructive, not destructive? And to link it to MST, we asked to add elements such as future tactile interaction, future target groups and using context and what is being communicated.}

Group composition also proved to influence speculative outcomes. The more homogeneous groups, such as haptics researchers in WS3, explored opportunities and threats in a more symmetrical way without privileging one over the other. One of the haptic experts of WS3 mentioned: ``\textit{I liked where we could have sincere discussion with haptic enthusiasts to create haptic content and future products. This part was also challenging since it is a field that is yet developed}.'' A more heterogeneous group composition, such as WS2 which included participants from law, ethics, and embodied practice, tended to arrive at more techno-critical perspectives. \textcolor{black}{For instance, discussions in WS2 frequently centred on consent, governance, and bodily autonomy, with domain experts framing these not simply as regulatory concerns but as challenges embedded in everyday social touch. Participants with backgrounds in ethics and law drew attention to how consent in social touch is often implicit and negotiated through contextual cues (P12), and to the legal complexities of establishing whether consent was given in a remote touch interaction (P10). These perspectives contrast with how consent has been discussed in prior expert-focused research \cite{barbareschi_moving_2025}, where it was primarily framed as a matter of user control and regulatory guidelines. This suggests that including perspectives beyond haptics expertise can reveal aspects of MST challenges that expert-only research may overlook.}

Although all workshops were held in person, we experimented with different material setups due to the different circumstances of the workshops, as WS3 was held at a large conference. In WS1 and WS2, participants worked on a shared digital canvas using laptops, while WS3 relied on posters, sticky notes, and physical materials. Digital tools supported legibility, easy revision, and reuse of workshop artefacts for analysis. Facilitators could quickly edit or clarify consequences when meanings were unclear, and participants could modify their ideas if desired. In WS3, the use of physical materials introduced different dynamics. Handwritten notes were occasionally difficult to read, both during the workshop and in later analysis. At the same time, using yarn and strings to connect and be able to re-connect consequences (see \autoref{fig:AnalogDigitalRipples}) – as we noticed is a dynamic and changing process – prompted longer and more engaged discussions about how effects were related. The act of cutting and pinning yarn became a playful moment. We noticed comparably longer discussions around interconnectedness through working with yarns as compared to the digital linking on Miro. Hence, the tangible engagement appeared to support more reflection on interdependence and systemic complexity, echoing recent HCI discussions and experiments on material form giving as a way to think with complex systems \cite{rosen_yarn_2025}.

Overall, our reflections position Future Ripples not as a fixed method but as a facilitation practice that requires ongoing attunement to framing, materials, group dynamics, and participation.

\section{Limitations and Future Work}
Our aim is not to claim an exhaustive account of all possible futures of MST, nor to generalise from the scenarios presented here. Rather, we offer these insights as provocations and pointers that can inspire further reflection and debate. We are aware that the future visions generated in our workshops are subjective, shaped by the participants’ own experiences, expertise, and perspectives. At the same time, we hope that the ideas articulated in this paper can serve as a starting point for haptic researchers and designers to critically engage with the distinctive challenges that may accompany the ongoing development of MST technologies. 

As noted in our methodological reflections, outcomes are shaped by participant group compositions, including individuals' professional backgrounds and lived experiences. \textcolor{black}{Furthermore, as each workshop used different what-if scenarios and allocated time differently across activities, we cannot fully separate the influence of group composition from these other factors. Future work could address this by conducting parallel workshops with different stakeholder groups speculating on the same scenario, allowing for more direct comparison of how expertise shapes speculative outcomes.} Our workshop design prioritised a progression from potential users through domain experts to technology designers, rather than breadth across all possible stakeholder groups. Future work could extend this enquiry to include stakeholders who would implement these technologies (e.g., industrial developers), practitioners who would use them in their practices (e.g., nurses, teachers, therapists, and somatic practitioners), and members of vulnerable communities who may be most affected by them.

For future workshops, we also see value in grounding futuring activities in more tangible experiences of MST. While our scenarios were based on discussion and imagination, the use of physical prototypes as speculative tools could provide a more embodied entry point into envisioning futures~\cite{wilde_2010_swing}. Bringing artefacts and prototypes into workshops may not only stimulate participants’ creativity but also elicit affective and bodily responses that are difficult to capture through verbal speculation alone (which is in line with our observations on using physical yarn to connect consequences). Such approaches could reveal new experiential dimensions of both opportunity and risk that remain hidden without a tactile or lived experience of the technology.



\section{Conclusion}

In this paper, we speculate on the possible impacts of Mediated Social Touch (MST) by engaging stakeholders ranging from potential users and domain experts to haptics researchers. Through a series of speculative workshops, participants expressed enthusiasm about the emerging opportunities MST might offer while also raising concerns about its possible risks. Key issues included adapting new ways of interaction, respecting individual boundaries, and addressing the ownership and manipulation of touch data. These reflections underscored the importance of ethical considerations in the design of MST systems and devices. Across the workshops, the imperative to establish safeguards that preserve the positive potential of MST consistently emerged—an idea aligned with the concept of ethical touch emphasised by~\citet{jewitt_2019_DigitalTouchEthics}.

We took a step further by adopting the RSR pipeline as an analytical lens to map the challenges and strategies that emerged from our participants' speculations~\cite{Taguchi_2023_Multichannel}. Through three distinctive challenges—encoded knowledge, complicated consent, and persistent touch—we demonstrated how the pipeline can help haptics researchers and developers identify where challenges emerge and how they might be considered at different stages of MST development. We also highlight the importance of complementary efforts beyond the technical and design domain, including policymaking and education.


Future work could extend this research by embedding the RSR pipeline into participatory design toolkits, investigating it across different cultural contexts, or prototyping consent-aware MST interfaces. In addition, these speculative insights could be translated into systematic design processes to inform emerging design principles or guidelines for MST. As haptic technologies continue to evolve, we hope our work supports more ethically grounded and socially responsive design practices. We also invite researchers to partake in the collaborative effort in shaping more preferable futures among the many plausible MST trajectories.

\begin{acks}
Wu was supported by the University of Sydney International Stipend Scholarship and the Taiwanese Government Scholarship to Study Abroad. Dr. Kim's participation was supported by the Electronics and Telecommunications Research Institute (ETRI) grant funded by the Korean government (Grant No. 26ZR1200 and 25YR1900). Dr. Withana is the recipient of an Australian Research Council Discovery Future Fellowship funded by the Australian Government. (Grant No. FT250100813) We thank all workshop facilitators and participants for their contributions to this research, and the IEEE World Haptics Conference 2025 workshop chairs and our speakers for their support in hosting our workshop at the conference. We also thank the anonymous DIS'26 reviewers and AC for their constructive feedback and suggestions on how to make this work stronger.
\end{acks}


\bibliographystyle{ACM-Reference-Format}
\bibliography{references, bibliography}

\appendix




\section{Participants' Background}

\begin{table*}[t]
\caption{Demographics and Background of Workshop Participants}
\Description{Table 3 presents demographic and background information for 24 participants across the three workshops. Each row includes participant number, gender, the workshop they attended (WS1 Medical, WS2 Social, or WS3 Educational/Emotional/Collaborational), their professional or academic background, experience in their field, and age group. Participants included a mix of students and researchers, such as veterinarians, biomedical researchers, interactive and design students, psychology students, senior HCI and haptics researchers, and law and cybersecurity researchers. Gender distribution was balanced across workshops, with both male and female participants represented. The table shows the diverse expertise and roles brought into the workshops, spanning medicine, design, psychology, law, data science, ethics, and haptics.}
\label{tab:participants}
\begin{tabular}{clllll}
\hline
\begin{tabular}[c]{@{}c@{}}Participants\\ Number\end{tabular} & \multicolumn{1}{c}{Gender} & \multicolumn{1}{c}{\begin{tabular}[c]{@{}c@{}}Workshop\\ (MST Purpose in What-if)\end{tabular}} & \multicolumn{1}{c}{Background} & \multicolumn{1}{c}{\begin{tabular}[c]{@{}c@{}}Experience in\\ their Field\end{tabular}} & \begin{tabular}[c]{@{}c@{}}Age \\ Group\end{tabular} \\ \hline
P1 & Female & WS1 (Medical Purpose) & Veterinarian Researcher & More than 10 years & 25-34 \\ \hline
P2 & Male & WS1 (Medical Purpose) & \begin{tabular}[t]{@{}l@{}}Interactive Design \\Master Student\end{tabular}  & 1-3 years & 18-24 \\ \hline
P3 & Male & WS1 (Medical Purpose) & Senior HCI Researcher & More than 10 years & 35-44 \\ \hline
P4 & Female & WS1 (Medical Purpose) & \begin{tabular}[t]{@{}l@{}}Digital Health \& Data Science\\Master Student\end{tabular} & 1-3 years & 25-34  \\ \hline
P5 & Male & WS1 (Medical Purpose) & Senior HCI Researcher & 4-6 years & 25-34 \\ \hline
P6 & Female & WS1 (Medical Purpose) & \begin{tabular}[t]{@{}l@{}}Interactive Design \\Master Student\end{tabular} & Less than 1 year & 25-34 \\ \hline
P7 & Female & WS2 (Social Purpose) & Psychology Student & 1-3 years & 18-24 \\ \hline
P8 & Female & WS2 (Social Purpose) & \begin{tabular}[t]{@{}l@{}}Dancer and Design \\Master Student\end{tabular}  & 1-3 years & 25-34 \\ \hline
P9 & Male & WS2 (Social Purpose) & Bio Medical Researcher & 1-3 years & 25-34 \\ \hline
P10 & \begin{tabular}[c]{@{}l@{}}Male\end{tabular} & \begin{tabular}[c]{@{}l@{}}WS2 (Social Purpose)\end{tabular} & \begin{tabular}[t]{@{}l@{}}Senior Law and Cybersecurity\\Researcher\end{tabular} & More than 10 years & \begin{tabular}[c]{@{}l@{}}55-64\end{tabular} \\ \hline
P11 & Female & WS2 (Social Purpose) & Psychology Master Student & 4-6 years & 25-34 \\ \hline
P12 & Male & WS2 (Social Purpose) & Ethic Researcher & More than 10 years & 25-34 \\ \hline
P13 & Male & WS3 (Educationanl Purpose) & Senior Haptics Researcher & More than 10 years & 45-54 \\ \hline
P14 & Male & WS3 (Educationanl Purpose) & Senior Haptics Researcher & More than 10 years & 45-54 \\ \hline
P15 & Female & WS3 (Educationanl Purpose) & Haptics Researcher & 1-3 years & 25-34 \\ \hline
P16 & Male & WS3 (Educationanl Purpose) & Haptics Researcher & 1-3 years & 18-24 \\ \hline
P17 & Male & WS3 (Emotional Purpose) & Haptics Researcher & 1-3 years & 25-34 \\ \hline
P18 & Female & WS3 (Emotional Purpose) & Senior Haptics Researcher & 6-10 years & 25-34 \\ \hline
P19 & Female & WS3 (Emotional Purpose) & Haptics Researcher & 1-3 years & 18-24 \\ \hline
P20 & Female & WS3 (Emotional Purpose) & Senior Haptics Researcher & 4-6 years & 18-24 \\ \hline
P21 & Male & WS3 (Collaborational Purpose) & Senior Haptics Researcher & More than 10 years & 45-54 \\ \hline
P22 & Male & WS3 (Collaborational Purpose) & Senior Haptics Researcher & More than 10 years & 55-64 \\ \hline
P23 & Female & WS3 (Collaborational Purpose) & Senior Haptics Researcher & 4-6 years & 35-44 \\ \hline
P24 & Male & WS3 (Collaborational Purpose) & Haptics Researcher & 1-3 years & 18-24 \\ \hline
\end{tabular}
\end{table*}

\cleardoublepage
\pagebreak
\section{Workshop Templates}

\begin{figure*}[b]
    \includegraphics[width=0.8\linewidth]{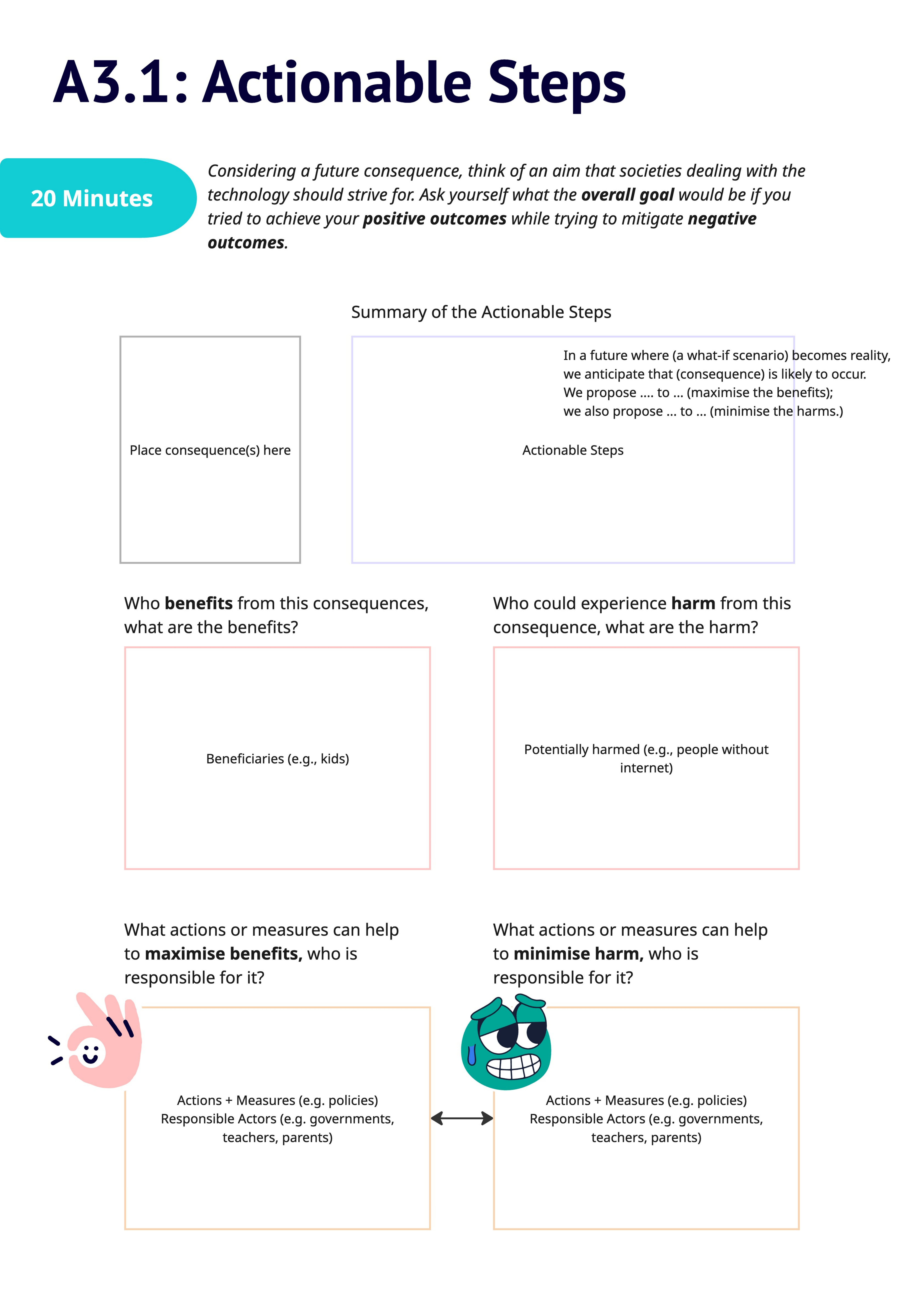}
    \captionsetup{justification=raggedright,singlelinecheck=false}
    \caption{A3.1: Actionable Steps Template}
    \Description{The figure shows a template used to guide participants in mapping future consequences and responses. Sections prompt users to identify consequences, beneficiaries, potential harms, and actions to maximise benefits or minimise harm, along with responsible actors. The template supports structured ethical reflection and actionable outcome generation. Note: Description generated with ChatGPT.}
    \label{fig:ActionableStepsTemplate}
\end{figure*}

\end{document}